\documentclass[%
superscriptaddress,
 amsmath,amssymb,
 aps,
prb,
twocolumn
]{revtex4-2}
\usepackage{graphicx}% Include figure files
\usepackage{dcolumn}% Align table columns on decimal point
\usepackage{bm}% bold math
\usepackage{diagbox}
\usepackage{hyperref}
\usepackage{braket}
\usepackage{dsfont}
\usepackage{amsfonts}
\usepackage{simplewick}
\usepackage{amsmath}
\graphicspath{{plots/}}

\begin{document}

\preprint{APS/123-QED}

\title{Study of the long-range transverse field Ising model with fermionic Gaussian states}% Force line breaks with \\

\author{Michael P. Kaicher}
\email[Correspondence to: ]{michael.p.kaicher(at)gmail.com}
\altaffiliation[Now at:  ]{BASF Digital Solutions, Next Generation Computing, Pfalzgrafenstr. 1, D-67056, Ludwigshafen, Germany }
\affiliation{Departamento de F\'isica Te\'orica, Universidad Complutense, 28040 Madrid, Spain}
%\affiliation{BASF SE, Quantum Chemistry, Carl-Bosch-Straße 38, D-67056, Ludwigshafen, Germany}

\author{Davide Vodola}
\altaffiliation[Now at:  ]{BASF Digital Solutions, Next Generation Computing, Pfalzgrafenstr. 1, D-67056, Ludwigshafen, Germany }
\affiliation{Dipartimento di Fisica e Astronomia, Universit\`a di Bologna, I-40129, Bologna, Italy}
\author{Simon B. J\"{a}ger}

\affiliation{Department of Physics and Research Center OPTIMAS, University of Kaiserslautern-Landau, D-67663 Kaiserslautern, Germany}

\date{\today}% It is always \today, today,

\begin{abstract}
We numerically study the one-dimensional long-range Transverse Field Ising Model (TFIM) in the antiferromagnetic (AFM) regime at zero temperature using Generalized Hartree-Fock (GHF) theory. The spin-spin interaction extends to all spins in the lattice and decays as $1/r^\alpha$, where $r$ denotes the distance between two spins and $\alpha$ is a tunable exponent. We map the spin operators to Majorana operators and approximate the ground state of the Hamiltonian with a Fermionic Gaussian State (FGS). Using this approximation, we calculate the ground state energy and the entanglement entropy which allows us to map the phase diagram for different values of $\alpha$. In addition, we compute the scaling behavior of the entanglement entropy with the system size to determine the central charge at criticality for the case of $\alpha>1$. For $\alpha<1$ we find a logarithmic divergence of the entanglement entropy even far away from the critical point, a feature of systems with long-range interactions.  We provide a detailed comparison of our results to outcomes of Density Matrix Renormalization Group (DMRG) and the Linked Cluster Expansion (LCE) methods. In particular, we find excellent agreement of GHF with DMRG and LCE in the weak long-range regime $\alpha\geq 1$, and qualitative agreement with DMRG in the strong-long range regime $\alpha \leq 1$. Our results highlight the power of the computationally efficient GHF method in simulating interacting quantum systems. 
\end{abstract}

\maketitle

%\tableofcontents

%-------------------------------------------------------
%-------------------------------------------------------
%-------------------------------------------------------
\section{Introduction\label{intro}}
Quantum phase transitions describe the behavior of quantum many-body systems at zero temperature when tuning a non-thermal control parameter, such as an applied magnetic field. The phase transition appears as a result of competing phases that describe the ground state at the corresponding parameter and typically lead to a fundamental change in the nature of the correlation present in the ground state. Quantum many-body systems can undergo a quantum phase transition and their study has lead to the discovery of many exotic collective phenomena such as superconducting ground states \cite{tinkham2004introduction}, long-range topological order \cite{haldane2017nobel}, and anyonic statistics\cite{stern2008anyons}.  Close to the critical point, the properties of many different physical systems can be classified by a universality class which is independent of the system size and only depends on the underlying dimensions and symmetries of the problem. In this situation, one can in many instances describe the many-body problem by an interacting spin system \cite{sachdev_2011}.

One of the paradigmatic microscopic models displaying a quantum phase transition is the Transverse Field Ising Model (TFIM) at zero temperature \cite{elliot1970ising}. This model is exactly solvable in the limit of short-range, nearest-neighbour interactions. However, the solution of this problem is much harder if one considers beyond nearest-neighbour or even long-range interactions \cite{koffel2012entanglement,knap2013probing,vodola2015long,fey2016critical,fey2019quantum}. Long-range interacting systems can host exotic states of quantum matter and are therefore of large scientific interest. Recent advances have made effective long-range spin-interactions experimentally accessible \cite{britton2012engineered,Schneider2012experimental,Friedenauer2008simulating,Jurcevic2014quasiparticle,bermudez2013dissipation}. In such systems, the effective interaction extends to all spins in the lattice and decays as a power law $1/r^\alpha$, where $r$ is the distance of the spins in the lattice and $\alpha$ is a  tunable algebraic exponent. In the experiments one can realize $0\leq\alpha\leq3$ which allows one to experimentally probe the regime of long-range interactions in spin systems \cite{britton2012engineered}.

In order to analyze the properties of a quantum many-body system, it is important to study large system sizes, which is in our case the number of spins $N$. The exponential scaling of the Hilbert space dimension with $N$ makes the ad-hoc diagonalization of such many-body problems illusive. Consequently, one demands numerical methods which are able to capture the qualitative behavior of the many-body system with a computational cost that displays a low scaling with $N$. A range of many-body methods of varying computational complexity have been applied to study finite size long-range quantum many-body systems, including Quantum Monte Carlo (QMC) \cite{humeniuk2016qunatum}, stochastic series expansion QMC \cite{koziol2021quantum}, a combination of QMC and renormalization group methods \cite{Laflorencie2005critical}, Lanczos exact diagonalization \cite{sandvik2010ground}, and Density Matrix Renormalization Group (DRMG) \cite{koffel2012entanglement,vodola2015long}. Recently, a method to study short-range quantum-lattice models in the thermodynamic limit, the Linked-Cluster Expansion (LCE), has been extended to allow for the study of long-range systems for $\alpha>1$ \cite{fey2016critical,fey2019quantum}.

In this work, we add Generalized Hartree-Fock (GHF) theory to this mix of methods. GHF is a \textit{mean-field method} which aims to approximate the ground state of an interacting quantum system as a free electron gas \cite{Bach1994}, where the latter describes a class of variational functions known as Fermionic Gaussian States (FGS). Due to its mean-field nature, GHF is a method with very low computational cost, where the most-demanding compute operation---the evaluation of the Pfaffian $\text{Pf}({\bf A})$ of a $M\times M$ matrix ${\bf A}$---scales at most as $\mathcal O(M^3)$ \cite{wimmer2012algorithm}.  Even though FGS describe ground or thermal states of quadratic fermionic Hamiltonians \cite{bravyi2004lagrangian}, they have been applied to various areas of quantum many-body physics with great success, most notably as ab-initio methods to obtain approximate ground states in electronic structure problems and to condensed matter systems~\cite{Bach1994,kraus2010generalized,shi2018variational}. In this paper, in order to find the FGS which best approximates the ground state of the long-range TFIM, we employ two physically-motivated methods which have been described in Ref.~\cite{kraus2010generalized}. The first one (ITE) derives the ground state using Imaginary Time Evolution. The second one (ZT) uses a self-consistent equation for the FGS ground  state covariance matrix. Using these  two methods we calculate the ground state energy and the entanglement entropy. By comparison of these results with the ones obtained from DMRG and ZT we will show that GHF is able to capture the qualitative and quantitative behavior of the long-range TFIM. This highlights the ability of GHF in predicting physically relevant material properties at computationally low cost.

This work is structured as follows. In Section~\ref{theory} we discuss the GHF theory which we then apply to the TFIM model described in ~\ref{ising_ham}. The introduced methods are used in Section~\ref{numerics} where we numerically study the ground state energy and the entanglement entropy. We conclude by summarizing our findings in Section~\ref{summary} and providing an outlook for future work.

%-------------------------------------------------------
%-------------------------------------------------------
%-------------------------------------------------------
\section{Theory\label{theory}}
%-------------------------------------------------------
\subsection{Long-range transverse field Ising model\label{ising_ham}}
In this work we consider the TFIM Hamiltonian describing a system of $N$ spins with open boundary conditions
	\begin{align}
		\hat{H}=\sum_{p=1}^N h_p\hat{\sigma}_p^z+\sum_{p< q}^NJ_{pq}\hat{\sigma}_p^x\hat{\sigma}_q^{x},\label{is50}
	\end{align}
where we introduced the transversal magnetic field strength $h_p=\cos(\theta)$, the interaction strength $J_{pq}=\sin(\theta)/|p-q|^\alpha$ and the Pauli matrices $\hat{\sigma}_p^a$ ($a\in\{x,y,z\})$ for each spin indexed by $p,q$. The magnetic field and interactions strengths are parameterized by the angle $\theta$ and the algebraic scaling of the interaction range is given by $\alpha$. In this work we furthermore focus on \textit{antiferromagnetic} (AFM) couplings which implies $J_{pq}>0$ or $\theta\in(0,\pi)$. Because the Hamiltonian is symmetric under the simultaneous transformations  
$\hat{\sigma}^z_p\mapsto-\hat{\sigma}^z_p$ and $\theta\to\pi-\theta$ we can restrict our study to $\theta\in(0,\pi/2]$.  

In a next step, we map the TFIM Hamiltonian onto a fermionic Hamiltonian. To this end, we use the Jordan-Wigner transformation
$\hat{\sigma}_p^+=\hat{c}_p^{\dag}e^{i\pi\sum_{q=1}^{p-1}\hat{c}_q^{\dag}\hat{c}_q}$ and $\hat{\sigma}_p^-=\hat{c}_p e^{-i\pi\sum_{q=1}^{p-1}\hat{c}_q^{\dag}\hat{c}_q}$ \cite{Jordan1928ueber}. Here, we used $\hat{\sigma}^{\pm}_p=[\hat{\sigma}^x_p\pm i\hat{\sigma}^y_p]/2$ and introduced the fermionic raising and lowering operators $\hat{c}_p^\dag$, $\hat{c}_p$, respectively. The latter obey the canonical anticommutation relations $\{\hat{c}_p,\hat{c}_q\}=0$ and $\{\hat{c}_p,\hat{c}_q^\dag\}=\delta_{p,q}$, where $\delta_{p,q}$ is the Kronecker delta and $\{\hat{A},\hat{B}\}=\hat{A}\hat{B}+\hat{B}\hat{A}$ denotes the anticommutator of two operators $\hat{A},\hat{B}$. Instead of analyzing the problem in the basis of the $2\times N$ fermionic operators $\hat{c}_p,\hat{c}_p^\dag$ we represent the Hamiltonian in $2N$ Majorana operators $\hat a_{2p-1}=\hat c_p^\dag + \hat c_p$ and $\hat a_{2p}=i(\hat c_p^\dag - \hat c_p)$. The latter posses the anticommutation relation $\{\hat a_l,\hat{a}_m\}=2\delta_{l,m}$ $(l,m=1,2,\dots,2N)$ and the Hamiltonian~\eqref{is50} in the Majorana representation is given by
\begin{align}
	\hat{H}=&-i\sum_{p=1}^{N}h_p\hat a_{2p-1}\hat a_{2p}+\sum_{p<q}^N(-i)^{q-p}J_{pq}\hat{a}_{2p}\hat{S}_{pq}\hat{a}_{2q-1}\label{is9}.
\end{align}
Here, we introduced the string operator $\hat{S}_{pq}=\prod_{k=p+1}^{q-1}\left(\hat{a}_{2k-1}\hat{a}_{2k}\right)$ which is the product of $2\times(q-p+1)$ Majorana operators.  For nearest-neighbour interactions, $\alpha=\infty$ and $J_{pq}=\delta_{p,q\pm1}$, this string operator becomes the identity, $\hat{S}_{pq}=1$ and $\hat{H}$ becomes quadratic in the Majorana operators. Consequently, the model can be described by free fermions and is therefore exactly solved by a FGS \cite{bravyi2004lagrangian}. In general, however, for the long-range TFIM we will need to include the contribution of the operator $\hat{S}_{pq}$. To avoid ambiguity, we use the term \textit{long-range} in this work for all systems with $\alpha<\infty$, since the spin interaction breaks up into a sum of terms proportional to $1/|p-q|^\alpha$, where  all lattice sites $p,q$ give non-zero contributions, and not just nearest-neighbour sites $p,p+1$ (as in the special case $\alpha\rightarrow \infty$). Often times, the term \textit{long-range} is reserved in literature for an algebraic exponent $\alpha=\sigma+d$ in a $d$-dimensional system for $\sigma<0$  (which in a one-dimensional system refers to the regime $\alpha<1$) \cite{fisher1972critical,defnu2015fixed}. Thus, to avoid confusion, in our work we will refer to $\alpha<1$ as the \textit{strong} long-range, and to $\alpha>1$ as the \textit{weak} long-range regime, while the special case $\alpha=1$ is marginal.

%-------------------------------------------------------
\subsection{Fermionic Gaussian States\label{fgs_intro}}
The formal definition of a FGS is given by \cite{bravyi2004lagrangian},
\begin{align}
    \hat\rho_{\text{GS}} =& \text{tr}\left(e^{-\beta \hat H_{\text{GS}}}\right)^{-1}e^{-\beta\hat H_{\text{GS}}},\label{a16}
\end{align}
where $\hat H_{\text{GS}} =\frac{i}{4}\hat{\bf a}^T{\bf G}\hat{\bf a}$ is a Hermitian operator, $\beta\in\mathds R$, $\hat{\bf a} = (\hat a_1,\hat a_2,\dots,\hat a_{2N})^T$ is a column vector of Majorana operators, and ${\bf G}$ is a $(2N\times 2N)$ real-valued and anti-symmetric matrix. FGS are fully described by the real and anti-symmetric covariance matrix $\boldsymbol{\Gamma}$ with entries
\begin{align}
\Gamma_{lm} = \frac{i}{2}\text{tr}\left(\hat \rho_{\text{GS}}[\hat a_l,\hat a_m]\right),\label{is12}
\end{align}
$l,m\in\{1,2,\dots,2N\}$, and where $[\hat A,\hat B]=\hat A\hat B-\hat B\hat A$ denotes the commutator of two operators $\hat A,\hat B$. While Eqs.~\eqref{a16}-\eqref{is12} describe both pure and mixed FGS, we only focus on pure FGS in this work, since we are interested in the ground state. Pure FGS are characterized by $\boldsymbol{\Gamma}^2=-\boldsymbol{1}_{2N}$ ($\boldsymbol{1}_k$ denotes the $(k\times k)$-identity matrix), and eigenvalues of the covariance matrix are given by $\lambda\in\{-1,1\}$. All information contained in the density matrix~\eqref{a16} of a FGS is also contained in its covariance matrix~\eqref{is12}. The expectation value of a single tensor product of Majorana or fermionic operators can be computed efficiently through Wick's theorem \cite{wick1950evalutation,bravyi2004lagrangian},
\begin{align}
    \text{tr}\left(\hat\rho_{\text{GS}}\hat a_{i_1}\hat a_{i_2}\cdots\hat a_{i_{2m}}\right) =&(-i)^m\text{Pf}\left(\left.\boldsymbol{\Gamma}\right|_{i_1i_2\dots i_{2m}}\right),\label{iis13}
\end{align}
where $i_1\neq i_2\neq \dots\neq i_{2m}$ for $i_k\in \{1,\dots,2N\}$ and $k=1,\dots,2N$. The matrix $\left.\boldsymbol{\Gamma}\right|_{i_1i_2\dots i_{2m}}$ denotes a $(2m\times 2m)$-submatrix of $\boldsymbol{\Gamma}$ with the corresponding rows and columns $i_1,i_2,\dots,i_{2m}$, and $\text{Pf}({\bf A})$ denotes the Pfaffian of a skew-symmetric matrix ${\bf A}$. 
%-------------------------------------------------------
\subsection{Approximating the ground state with a fermionic Gaussian State\label{fgs_minimization}}
Using Wick's theorem~\eqref{iis13}, we are able to compute the energy expectation value \begin{align}
E(\boldsymbol{\Gamma})=&\text{tr}\left(\hat \rho_{\text{GS}}\hat H\right)\label{eq:Energy},
\end{align}
which results in  
\begin{align}
E(\boldsymbol{\Gamma})
=& -\sum_{p=1}^N \frac{h_p}{2}\left( \Gamma_{2p-1,2p}-\Gamma_{2p,2p-1}\right)\nonumber\\&+\sum_{p<q}^{N}J_{pq}(-1)^{q-p}\text{Pf}\left\{\left.\Gamma\right|_{2p,2p+1,\dots,2q-1}\right\},\label{is27}
\end{align}
for the Hamiltonian given by Eq.~\eqref{is50}.
In order to approximate the ground state of the Hamiltonian within the family of FGS, one has to find a covariance matrix $\boldsymbol{\Gamma}$ which minimizes $E(\boldsymbol{\Gamma})$. While one can apply any constrained optimization method, in the following, we will discuss two particular algorithms for finding the optimal $\boldsymbol{\Gamma}$, which we will use in Section~\ref{numerics}.

\paragraph{Imaginary Time Evolution (ITE)}
The first algorithm performs an Imaginary Time Evolution (ITE) under the constraint that Wick's theorem holds throughout the evolution. This constraint guarantees that the evolved state remains a FGS, and leads to an equation of motion for the corresponding covariance matrix, 
\begin{align}
    \frac{d\boldsymbol{\Gamma}}{d\tau} = \frac{1}{2}[\boldsymbol{\Gamma},[\boldsymbol{\Gamma},{\bf H}^{(\text{mf})}]],\label{is25}
\end{align}
where $\tau\in\mathds R$ denotes the imaginary time. We derive Eq.~\eqref{is25} in Appendix~\ref{derivation_ITE}. The central quantity is hereby the mean-field Hamiltonian ${\bf H}^{(\text{mf})}(\boldsymbol{\Gamma})$ which is the gradient of the energy with respect to the covariance matrix,  
\begin{align}
    H^{(\text{mf})}_{lm}=4\frac{dE(\boldsymbol{\Gamma})}{d\Gamma_{lm}}.\label{eq:mf}
\end{align}
This term can be computed explicitly by using identities for the matrix derivative of a Pfaffian, which we have also employed in Appendix~\ref{App:Wicktheorem}.

We solve Eq.~\eqref{is25} iteratively, by discretizing the ITE into small time steps $\Delta\tau$. Starting from a random initial covariance matrix, we  evolve the covariance matrix through $\boldsymbol{\Gamma}(\tau+\Delta\tau) \approx {\bf O}(\Delta\tau)\boldsymbol{\Gamma}(\tau) {\bf O}(\Delta\tau)^T$, where  ${\bf O}(\Delta\tau) = e^{\frac{1}{2}\left[{\bf H}^{(\text{mf})},\boldsymbol{\Gamma}\right]\Delta\tau}$ is an orthogonal matrix. As explicitly shown in Ref.~\cite{kraus2010generalized}, this approach preserves the purity of the FGS, while ensuring a monotonic decrease of the energy in each iteration. 

\paragraph{Zero Temperature (ZT)}
The second algorithm uses a self-consistent equation for the steady-state solution of Eq.~\eqref{is25}.
In this algorithm, for a given $\boldsymbol{\Gamma}$, we diagonalize the mean-field matrix  $i{\bf H}^{(\text{mf})}={\bf U}{\bf D}{\bf U}^\dag$ and recalculate
\begin{align}
    \boldsymbol{\Gamma} = i {\bf U}\text{sgn}({\bf D}){\bf U}^\dag.\label{is39}
\end{align}
Here, ${\bf U}$ is a unitary matrix, ${\bf D}$ is a diagonal matrix containing the real eigenvalues of $i{\bf H}^{(\text{mf})}$, and $\text{sgn}({\bf D})$ is the sign function applied to the diagonal entries of ${\bf D}$. From this $\boldsymbol{\Gamma}$ we recalculate $i{\bf H}^{(\text{mf})}$ and repeat the procedure until the covariance matrix is converged. One can check that the solution of this algorithm is also a stationary state of Eq.~\eqref{is25} with $\boldsymbol{\Gamma}^2=-{\bf 1}$.

In both algorithms we choose several random initial covariance matrices $\boldsymbol{\Gamma}_{\text{init}}$ to ensure unbiased results. A random $\boldsymbol{\Gamma}_{\text{init}}$ is generated through $\boldsymbol{\Gamma}_{\text{init}}={\bf O}^T\boldsymbol{\Omega} {\bf O}$, where ${\bf O}$ is a random orthogonal matrix and we defined the block diagonal matrix $\boldsymbol{\Omega}= \bigoplus_{k=1}^N(-1)^{r_k}\left(\begin{smallmatrix}0&1\\-1&0\end{smallmatrix}\right)$, where $r_k\in\{0,1\}$ is chosen randomly and $\bigoplus$ denotes the direct sum. After convergence of the corresponding algorithm we achieve a stationary solution $\boldsymbol{\Gamma}_{\text{st}}$. With the help of this solution we can then find the GHF approximation to the ground state energy given by $E({\boldsymbol{\Gamma}_{\text{st}}})$. Besides the energy and entanglement entropy introduced in the following section, the covariance matrix also allow us direct access to quantum correlations.
%-------------------------------------------------------
\subsection{Entanglement entropy and central charge\label{fgs_entglentr}}
Entanglement entropy is a well-studied measure for the amount of quantum correlations in a pure quantum state~\cite{amico2008entanglement, horodecki2009quantum}. It is defined as $S_{N_{\mathcal A}} = -\text{tr}(\hat \rho_{\mathcal A}\log(\hat \rho_{\mathcal A}))$, where $\mathcal A$ describes a subsystem containing $N_{\mathcal A}$ spins. The reduced density matrix $\hat \rho_{\mathcal A}=\text{tr}_{\mathcal B}(\hat \rho)$ is obtained by performing a partial trace over the disjoint subsystem $\mathcal B$, with $N_{\mathcal B} = N - N_{\mathcal A}$ spins. %We reserve $S_{\mathcal A}=S_{N/2}$ for the special case where $\mathcal A$ contains the $N_{\mathcal A}=N/2$ leftmost spins. 
For the spins numbered as $\mathcal{A}=\{1,2,\dots,N/2\}$ we define the corresponding Majorana operators by $\mathfrak M_{\mathcal A}=\{1,2,\dots,N-1,N\}$. The entanglement entropy is then fully determined by the matrix $\boldsymbol{\Gamma}_{\mathcal A}=\left.\boldsymbol{\Gamma}\right|_{\mathfrak M_{\mathcal A}}$ and can be calculated with \cite{peschel2003calculation, vidal2003entanglement, kraus2010generalized}
\begin{align}
S_{N/2}=& \frac{N}{2}\log(2) -\frac{1}{2}\text{tr}\left[\left({\bf 1}_{N}+i\boldsymbol{\Gamma}_{\mathcal A}\right)\log\left({\bf 1}_{N}+i\boldsymbol{\Gamma}_{\mathcal A}\right)\right].\label{as33}
\end{align}

For short-range 1D systems, the entanglement entropy typically follows two different scalings: for gapped phases, $S_{N/2}$ saturates to a constant value independent of $N$ and thus obeys the so-called area law~\cite{eisert2010colloquium}. For gapless phases, the entanglement entropy exhibits the following behavior~\cite{Calabrese2004entanglement}
\begin{align}
S_{N/2}=\frac{c}{6}\log(N)+B, \label{eq:EE}
\end{align}
where $c$ is the central charge characterizing the universality class of the system and $B$ is a non-universal constant. For the nearest-neighbour TFIM at $\alpha=\infty$ the value of $c=1/2$ can be found exactly.

For long-range systems we need to differentiate between \textit{weak} long-range interactions, $\alpha>d=1$, and the \textit{strong} long-range interactions, $\alpha<d=1$.

For \textit{weak} long-range interactions and a non-vanishing energy gap we expect also an area law scaling, implying that $S_{N/2}$ is independent of $N$. For the case of a vanishing gap one also finds a logarithmic divergence~\cite{Jin2004quantum,Its2005entanglement,keating2005entanglement,eisert2010colloquium} following Eq.~\eqref{eq:EE}.

For \textit{strong} long-range interactions in the AFM-TFIM we expect instead a logarithmic divergence of the entanglement entropy, where $S_{N/2}$ obeys Eq.~\eqref{eq:EE} and one can find $c\neq0$ even in presence of a non-vanishing gap~\cite{eisert2006general,mueller2012anomalous, vodola2014kitaev, ares2015entanglement}. In this regime $c$ is strictly speaking not a central charge but because of the same functional dependence of $S_{N/2}$ in Eq.~\eqref{eq:EE}, we also denote $c$ as the effective central charge.

\section{Results\label{numerics}}
%-------------------------------------------------------
\subsection{Phase diagram\label{phase_diagram}}
In this section, we show that a computationally inexpensive GHF mean-field approach can reproduce the phase diagram of the AFM-TFIM for a wide range of values $\alpha$, both in the weak and strong long-range regime, and is able to locate the point of the phase transition for $\alpha\geq 1$ in excellent agreement with  state-of-the-art numerical methods.

\begin{figure}[t]
\center
\includegraphics[width=1\columnwidth]{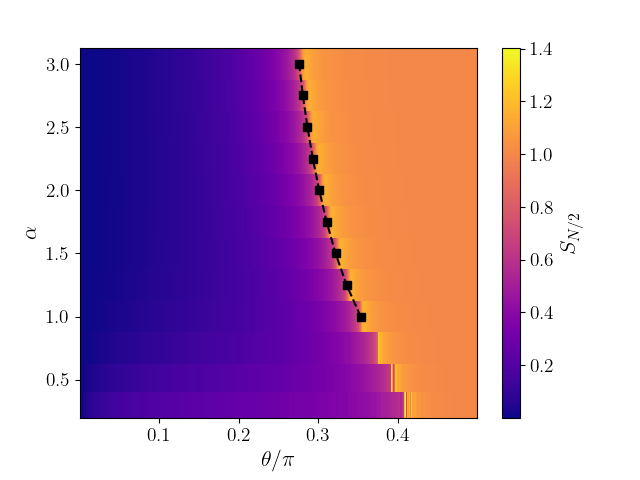}
\caption{\label{Fig:1}  We plot the entanglement entropy $S_{N/2}$ from the covariance matrix obtained through the ZT algorithm for a system size $N=100$, $\alpha \in\{0.30,0.50,0.75,1.00,\dots,3.00\}$, and $\theta\in(0,\pi/2)$. Black squares represent the quantum critical points $\theta_c^\infty/\pi$ in the thermodynamic limit, which are listed in Tab.~\ref{Tab:1}, while the dashed line serves as a guide to the eye. }
\end{figure}

As a first benchmark and in the same spirit of Ref.~\cite{koffel2012entanglement} we map the phase diagram by calculating the entanglement entropy for a wide range of values of $\alpha$, from \textit{weak} to \textit{strong} long-range interactions, and for $\theta\in(0,\pi/2)$. The values of $S_{N/2}$ [Eq.~\eqref{as33}] computed with the ZT GHF method are visible in Fig.~\ref{Fig:1} for $N=100$.
For $\theta=0$ the interactions vanish and $S_{N/2}=0$ for all values of $\alpha$. This represents the phase where all spins are uncorrelated and align with the external magnetic field. However, when $\theta$ and therefore the AFM interactions are increased, the minimization of the interaction energy competes with the external magnetic field. This is accompanied by an increase of $S_{N/2}$. Dependent on $\alpha$, there is a critical value $\theta_c(\alpha)$ beyond which the spins favor an AFM order. This transition is highlighted in Fig.~\ref{Fig:1} by a sharp rise of $S_{N/2}$.  Our findings are in qualitative agreement with the ones obtained in Ref.~\cite{koffel2012entanglement} from DMRG calculations. To compare our results also quantitatively, we will now focus on the \textit{weak} and \textit{strong} long-range interactions cases separately.

\subsection{Weak long-range interactions}

\subsubsection{Comparison of GHF and DMRG}

\begin{figure}[h!]
\flushleft(a)\\
\center\includegraphics[width=1\columnwidth]{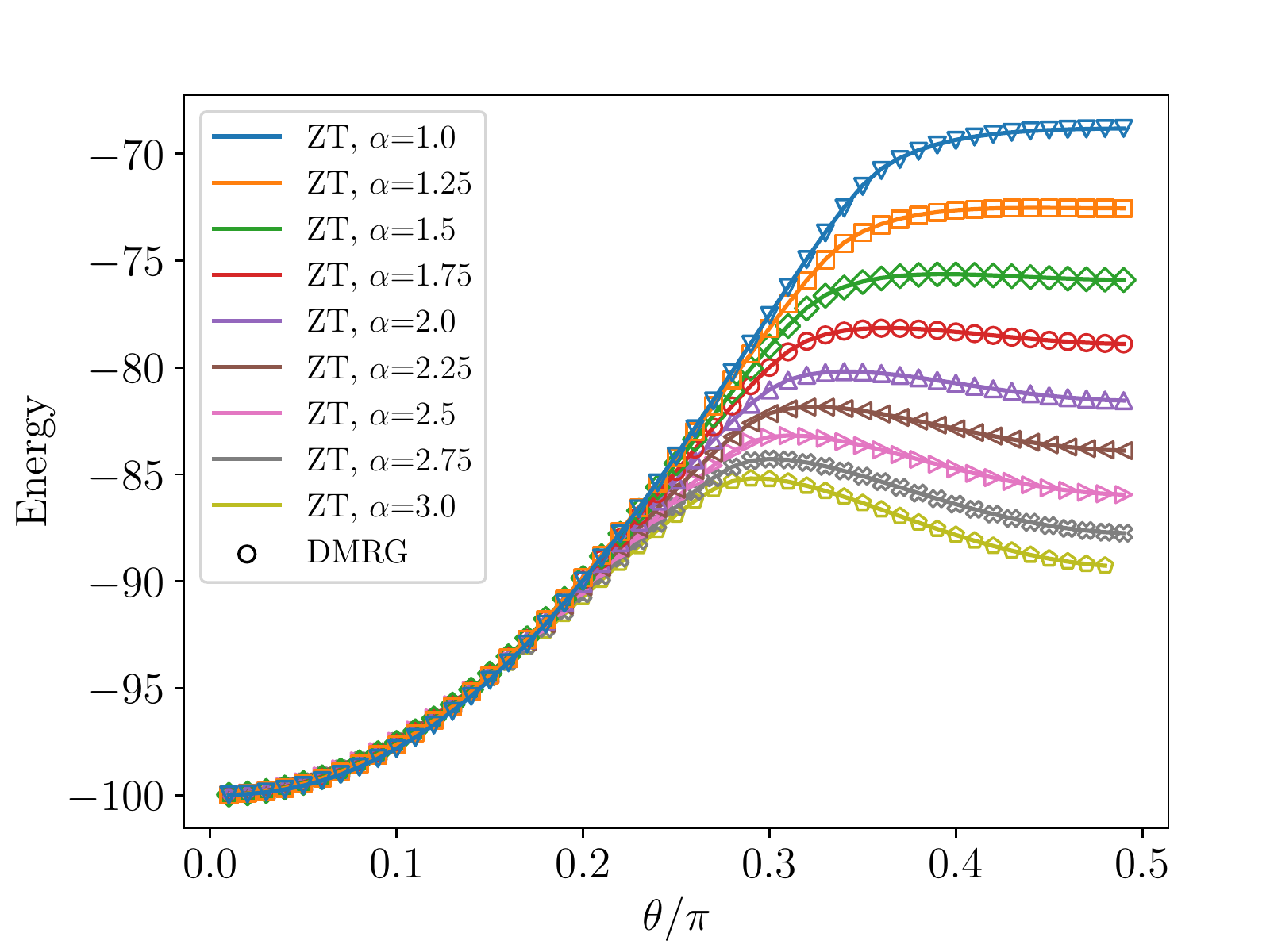}
\flushleft(b)\\
\center\includegraphics[width=1\columnwidth]{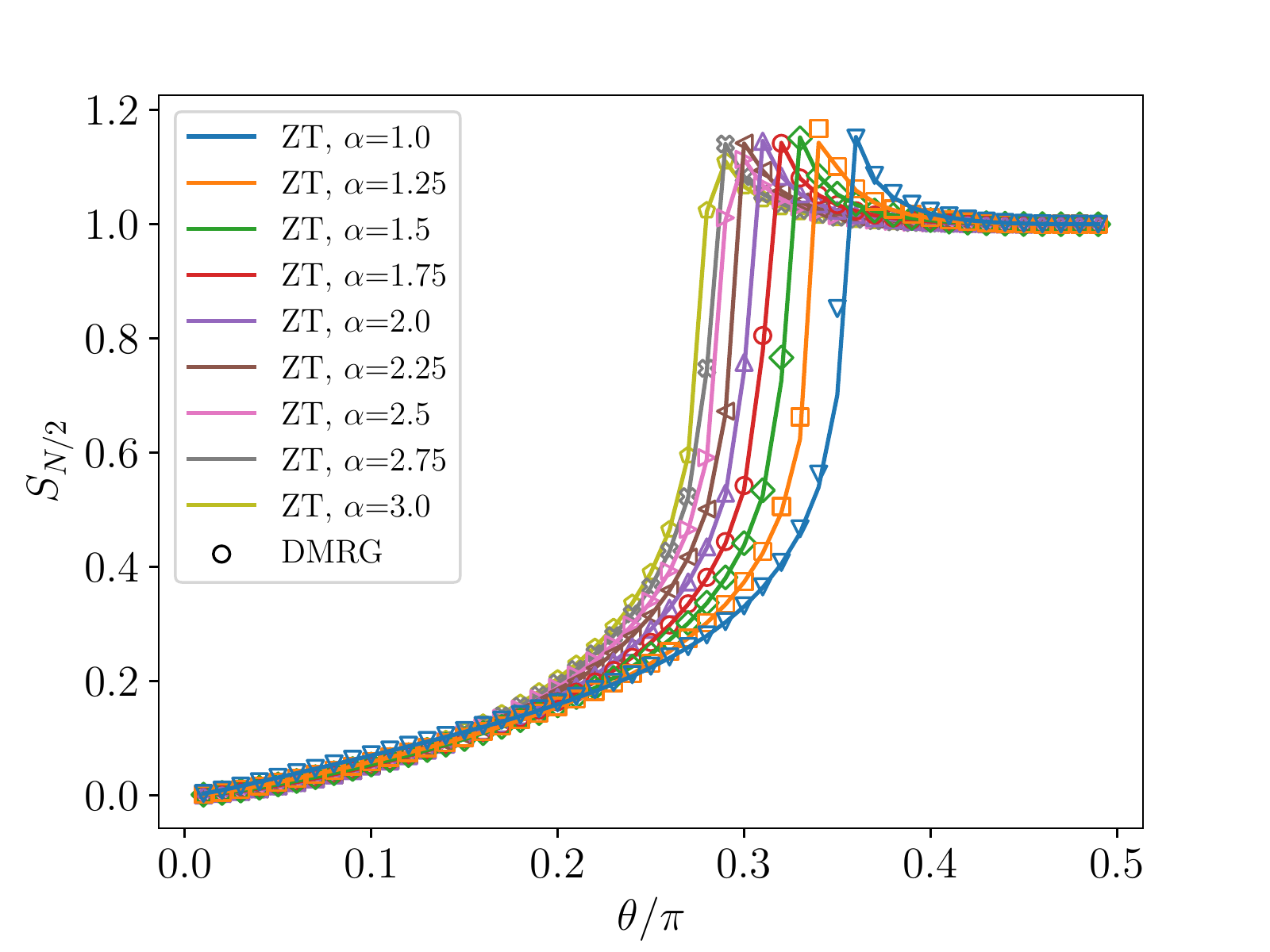}%
\caption{\label{Fig:2} For a system of size $N=100$ and exponents $\alpha\in[1,3]$, we plot (a) the energy $E$   and (b) the entanglement entropy $S_{N/2}$ (bottom), as defined in Eqs.~\eqref{eq:Energy} and ~\eqref{as33}, obtained from the covariance matrix of the ZT algorithm (solid lines) and compare it to DMRG (hollow markers). }
\end{figure}

For $\textit{weak}$ long-range interactions, $\alpha\geq 1$, we show the ground state energy and the entanglement entropy in Fig.~\ref{Fig:2}(a) and Fig.~\ref{Fig:2}(b), respectively. 

The solid lines represent the results obtained from the GHF theory while hollow markers represent the results obtained from DMRG simulations. Both simulation methods predict a rather smooth behavior of the energy in Fig.~\ref{Fig:2}(a). For larger values of $\alpha\geq 1.5$ we find a maximum and a decrease beyond the maximum point. The GHF and DMRG simulations agree perfectly.

The entanglement entropy, visible in Fig.~\ref{Fig:2}(b), shows for all values and both simulations methods a very quick increase and a pronounced singularity. The latter is an indicator for the phase transition point. Beyond this point we find again a decrease of the entanglement entropy. Both methods, GHF and DMRG, are in very good agreement. 

\subsubsection{Threshold and central charge}

\begin{figure}
    \center
    \includegraphics[width=1\columnwidth]{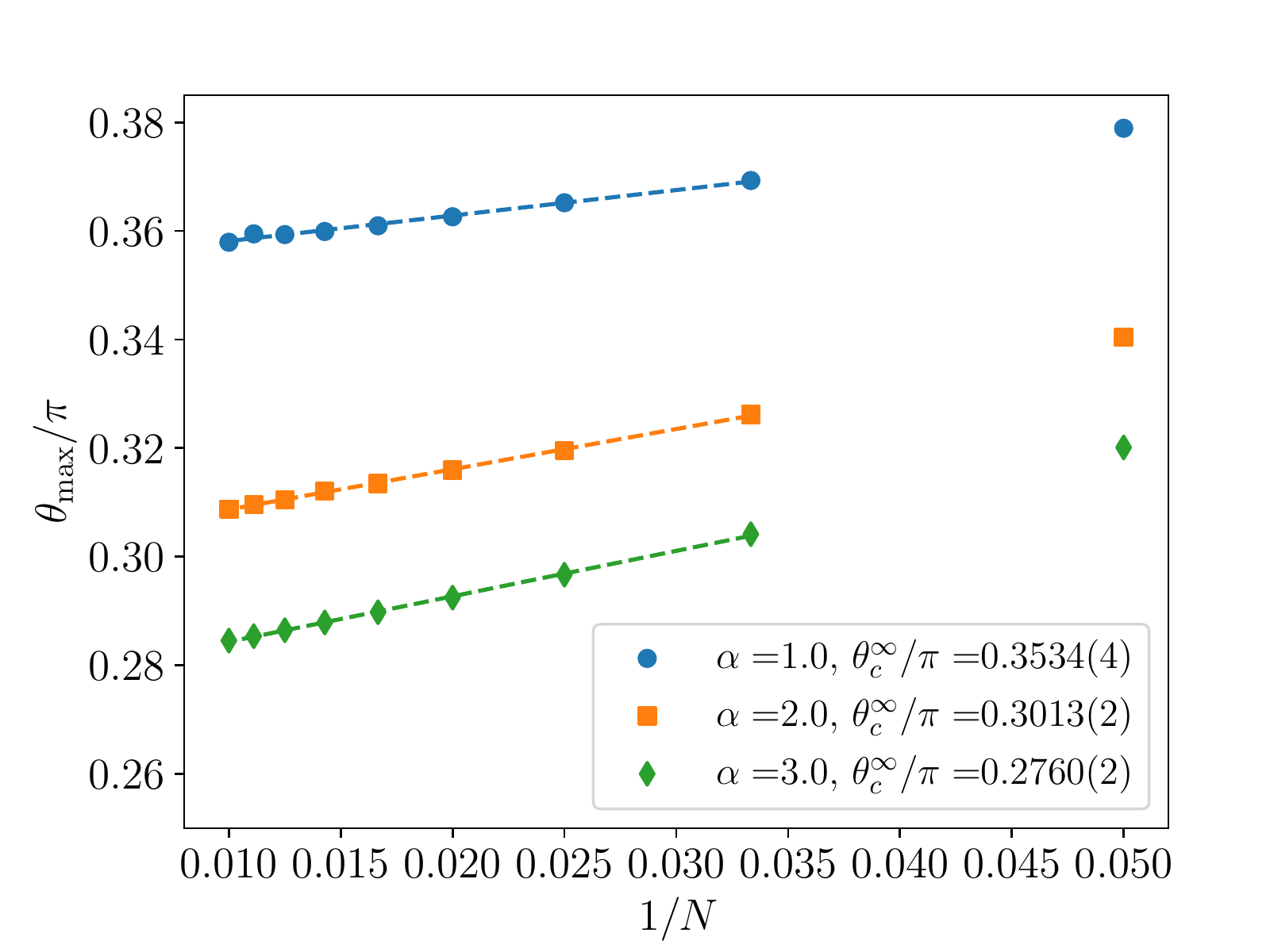}
    \caption{\label{Fig:3} Example for the fit of Eq.~\eqref{eq:threshfit} to the value of $\theta_\mathrm{max}$ obtained by maximization of the entanglement entropy $S$ with FGS. The thresholds $\theta_c^\infty/\pi$ are shown for the respective cases $\alpha\in\{1,2,3\}$, see Tab.~\ref{Tab:1} for more details.}
\end{figure}

 In order to find a value for the threshold at $N\to\infty$, we are performing a finite-size scaling. For this we carry out analogue simulations for a range of smaller system sizes $N\in\{20,30,\dots,100\}$. Then we find numerically the maximum of the entanglement entropy of the half chain $S_\mathrm{max}=S_{N/2}(\theta_{\mathrm{max}})$ and the corresponding value $\theta_{\mathrm{max}}$. The latter is found using the optimizer \texttt{scipy.optimize.fminbound()}  which is pre-implemented in python. For every value $\theta$ examined by the optimizer we find the optimal FGS for the corresponding Hamiltonian. Optimizing $S_{N/2}$ over $\theta$ can be achieved as FGS provide a way for calculating $S_{N/2}$ polynomially in $N$, see Eq.~\eqref{as33}.  We then use the following finite-size scaling law~\cite{nishimoto2011tomonaga}
 \begin{align}
 \theta_{\mathrm{max}}(N)=\theta_c^\infty+\frac{a}{N}, \label{eq:threshfit}
 \end{align}
 where $\theta_c^\infty$ is the threshold at $N\to\infty$ and $a$ is a fitting parameter which determines the finite-size scaling. Fitting Eq.~\eqref{eq:threshfit} to the numerically obtained data of $\theta_{\mathrm{max}}$ reveals the $\theta_c^\infty$ in the thermodynamic limit. In Fig.~\ref{Fig:3} we provide examples for the fits that are used to calculate $\theta_c^\infty$.
 We perform these fits for various values of $\alpha$ and the results for the threshold are collected in Tab.~\ref{Tab:1}.
 \begin{table}[b]
\center
\begin{tabular}{|c|llll|}
\hline
\diagbox{$\alpha$}{$\theta_{c}^{\infty}/\pi$} &FGS& LCE & DMRG & DMRG* \\
\hline
1.00&0.3534(4)&-& 0.3509& -\\
1.25& 0.3357(1)& 0.35(5)& -& -\\
1.50& 0.3218(1)& 0.3213(5)& 0.3226& -\\
1.75& 0.3106(1)& -& -& -\\
2.00& 0.3013(2)& 0.3026(8)& 0.3027& 0.3021\\
2.25& 0.2932(2)& 0.294(4)& -& -\\
2.50& 0.2865(1)& 0.2871(11)& -& -\\
2.75&0.2807(2)& -& -& -\\
3.00& 0.2760(2)& 0.27722(25)& 0.2782& -\\
\hline
\end{tabular}
\caption{\label{Tab:1}%
The critical points $\theta_c^{\infty}/\pi$ obtained from Eq.~\eqref{eq:threshfit} with FGS and ZT, in comparison to LCE \cite{fey2019quantum}, and DMRG \cite{koffel2012entanglement}, DMRG*\cite{vodola2015long} results. The values are obtained for various exponents $\alpha$ and for simulations up to $N=100$ spins. The error indicated in the FGS column in round brackets is the standard deviation for the intersect of a linear regression fit of $\theta_{\text{max}}/\pi$ as a function of $1/N$.
}
\end{table}
In addition, we have plotted the results of $\theta_c^\mathrm{max}$ in Fig~\ref{Fig:1} as black squares which mark the sudden spike of the entanglement entropy.  In Tab.~\ref{Tab:1} we compare the results obtained from the GHF theory with the ones obtained from LCE calculations~\cite{fey2019quantum}, DMRG data of Ref.~\cite{koffel2012entanglement} (labeled DMRG) and Ref.~\cite{vodola2015long} (labeled DMRG*). We find in general very good agreement of the thresholds obtained from the different methods. 

\begin{figure}
    \flushleft(a)\\
    \center
    \includegraphics[width=1\columnwidth]{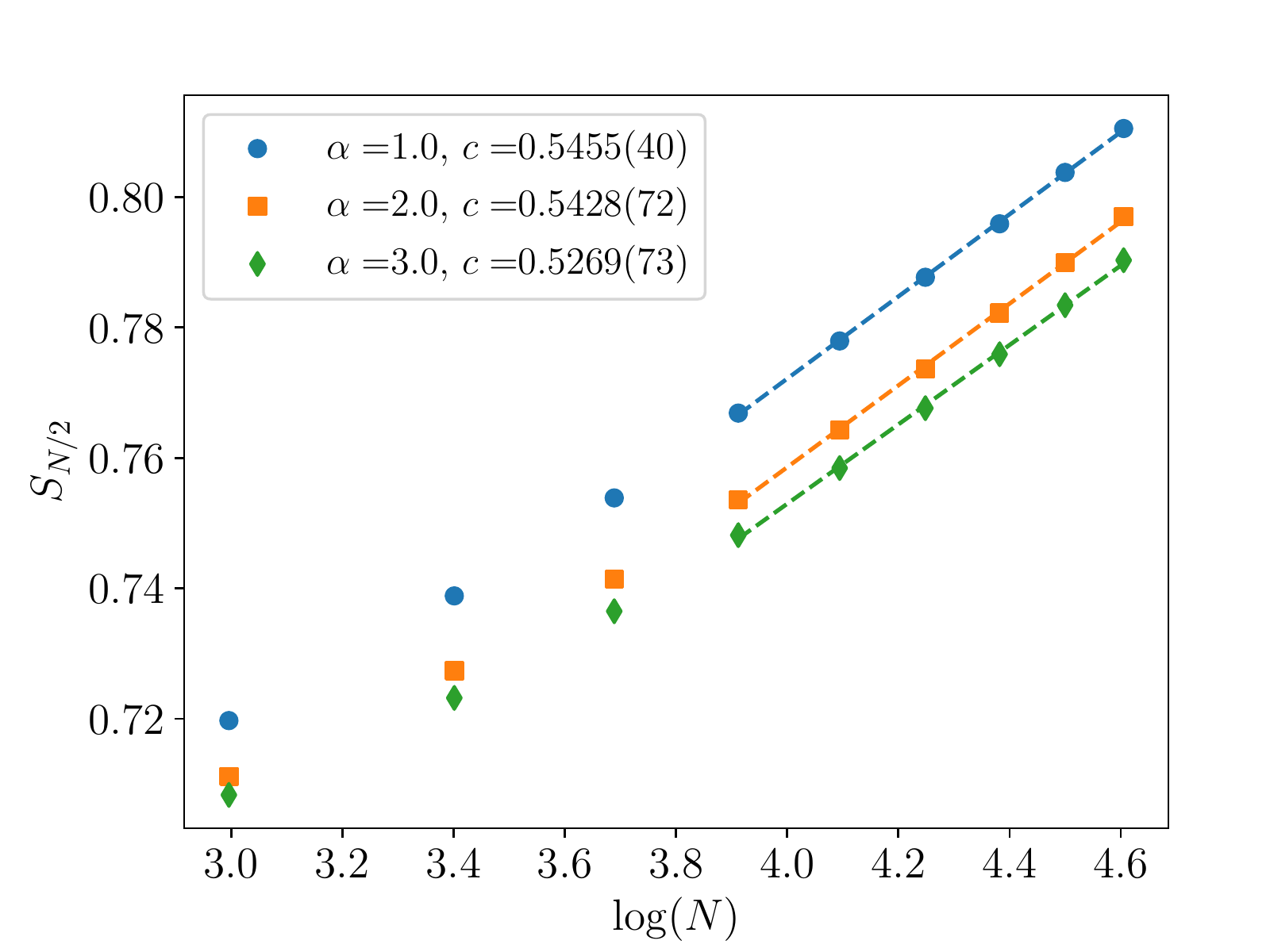}
    \flushleft(b)\\
    \includegraphics[width=1\columnwidth]{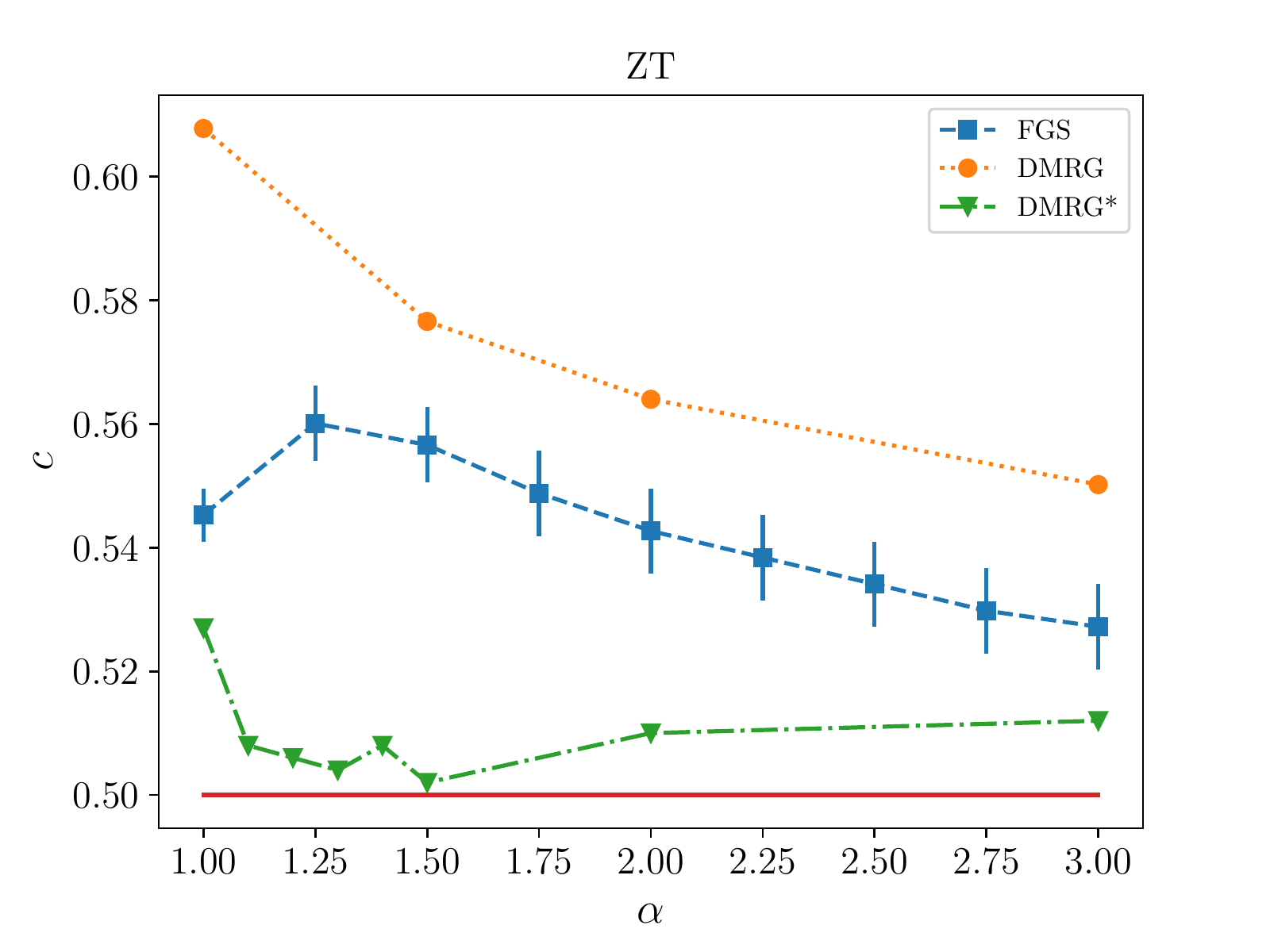}
    \caption{\label{Fig:4} (a) Extracting the central charge. Using the ZT algorithm for various $\alpha$, here exemplified by $\alpha\in\{1,2,3\}$, we  plot  the entanglement entropy $S_{N/2}$ against $\log(N)$. For each $\alpha$ we perform a linear regression fit, neglecting the system sizes $N\in\{20,30,40\}$ to mitigate finite size effects. (b) Central charge $c$ obtained from finite-size scaling up to system size $N=100$ of FGS evolutions through the ZT algorithm (blue squares) for the AFM long-range TFIM. For comparison, DMRG results from finite-size scaling of system sizes of up to $N=100$ from Ref.~\cite{koffel2012entanglement} ('DMRG', orange square) and ~\cite{vodola2015long} ('DMRG*', green triangles) are included. The red horizontal line represents the value $c=1/2$ which describes the Ising universality class. Error bars represent the standard deviation from the linear regression fit. }
\end{figure}

Besides the threshold $\theta^\infty_c$ we can also extract the scaling of the maximum entropy $S_\mathrm{max}=S(\theta_\mathrm{max})$. At the critical point we use the scaling law~\cite{Calabrese2004entanglement} given by Eq.~\eqref{eq:EE}. We fit Eq.~\eqref{eq:EE} to the maximum values $S_\mathrm{max}$ as displayed in Fig.~\ref{Fig:4}(a).
From these fits we extract the central charge $c$, which is shown in Fig.~\ref{Fig:4}(b) as function of $\alpha$. The central charge is always above the result $c=1/2$ expected from the short-range TFIM. We also compare our results to different DMRG results of Ref.~\cite{koffel2012entanglement,vodola2015long}.  We find that the central charges obtained from FGS are systematically smaller than the values provided by Ref.~\cite{koffel2012entanglement} and larger than the DMRG results of Ref.~\cite{vodola2015long}. The central charge $c$ is monotonically decreasing in the weak long-range regime, but drops at the onset of the strong long-range regime at $\alpha=1$. In conclusion, we found that the results of the GHF method are in good qualitative and quantitative agreement with state-of-the-art numerical methods for $\textit{weak}$ long-range interactions.

\subsection{Strong long-range interactions}

\subsubsection{Comparison of GHF and DMRG}
\begin{figure}[t]
    \flushleft(a)\\
\center\includegraphics[width=1\columnwidth]{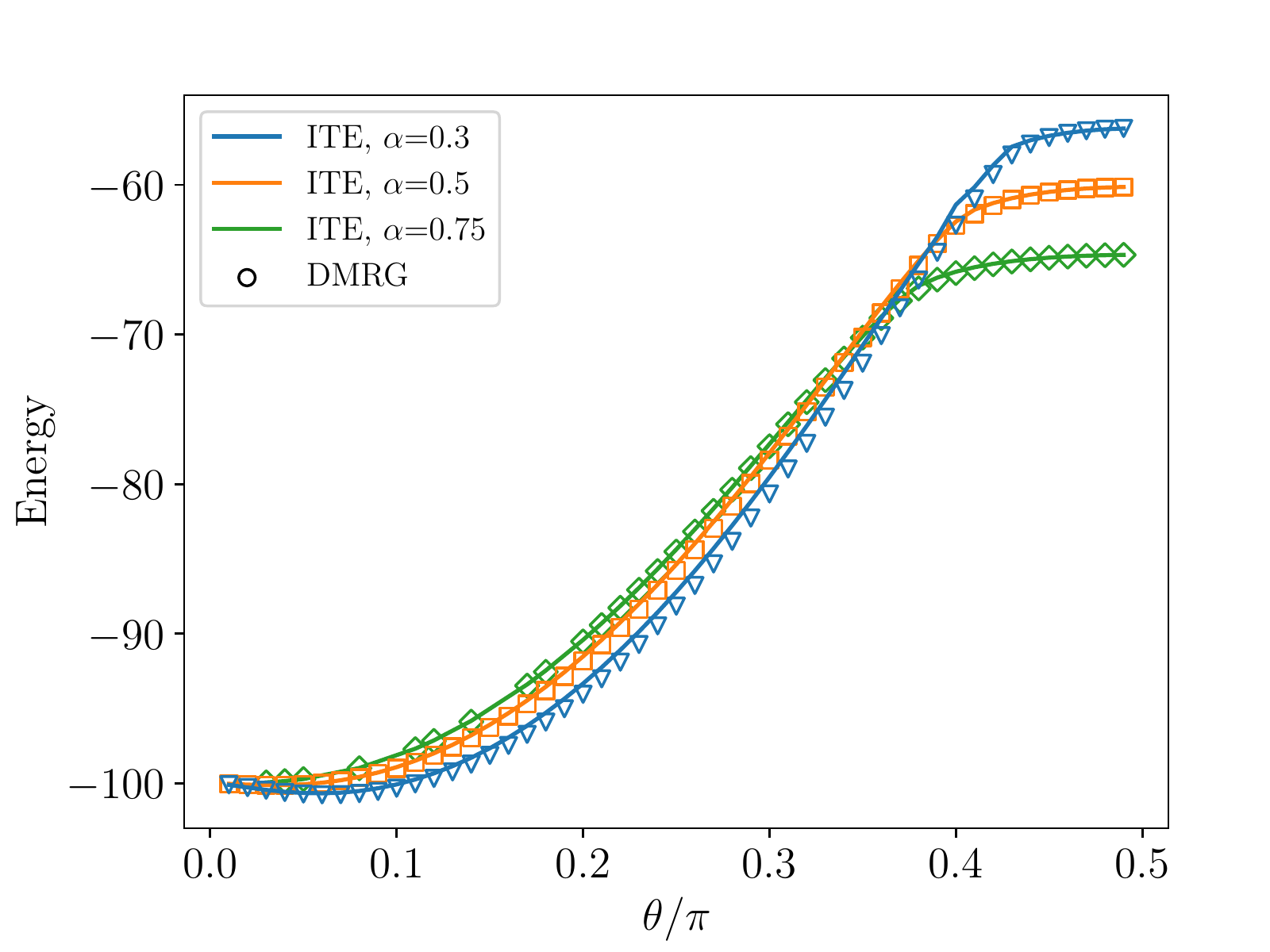}
\flushleft(b)\\
\center	\includegraphics[width=1\columnwidth]{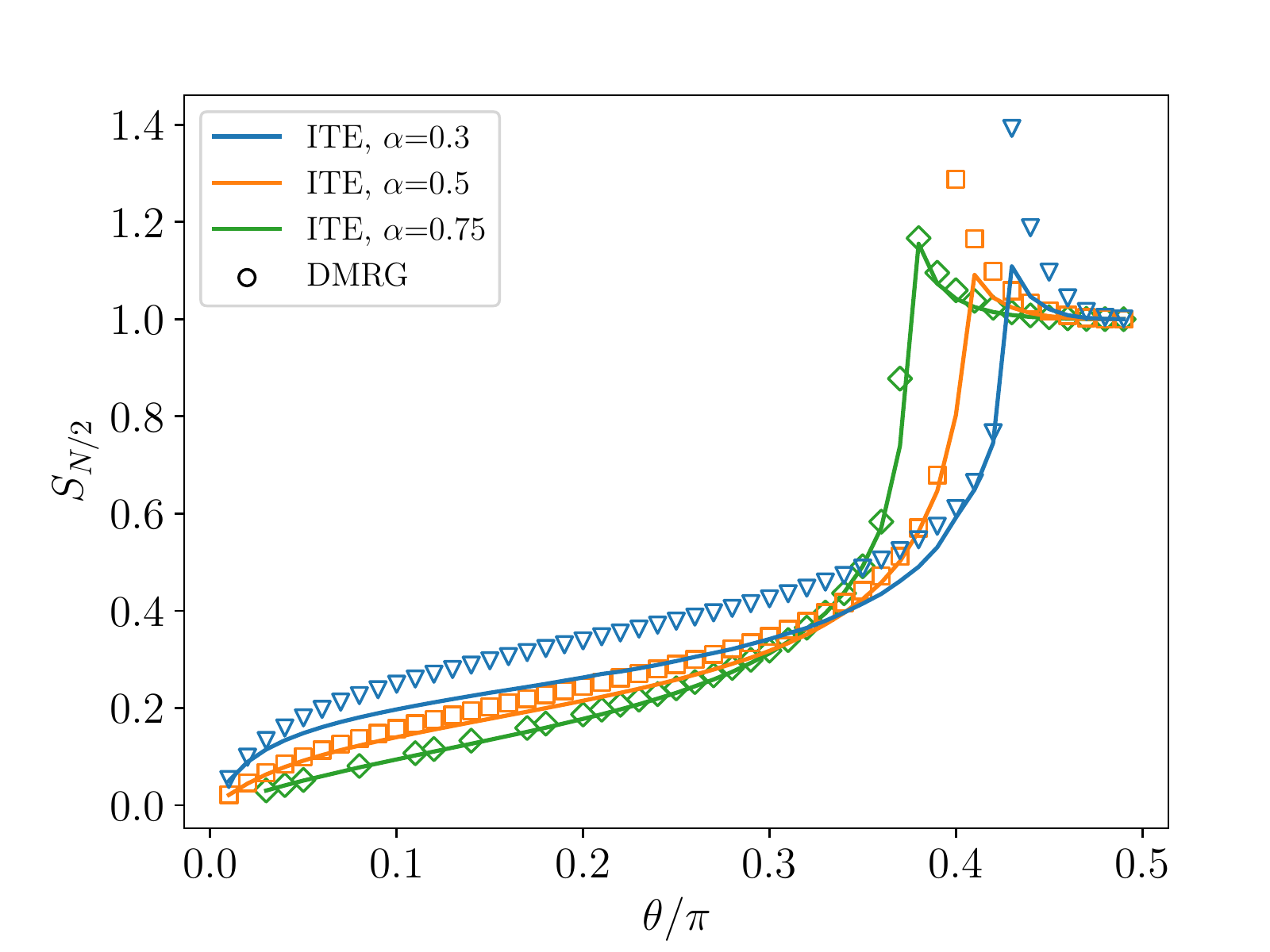}
    \caption{\label{Fig:5} (a) Energy and (b) entanglement entropy obtained from the covariance matrix of the ITE algorithm (solid lines) and DMRG (empty markers) simulations for $N=100$ and $\alpha\in\{0.3,0.5,0.75\}$. }
\end{figure}

We will now shift our focus to the regime of \textit{strong} long-range interactions, $\alpha<1$. We first plot the ground state energy and the entanglement entropy in Fig.~\ref{Fig:5}(a) and Fig.~\ref{Fig:5}(b) for three different values of $\alpha<1$ of size $N=100$.
In Fig.~\ref{Fig:5}(a) we obtain for all three values of $\alpha$ a monotonously increasing energy with $\theta$. This is different to the case of $\textit{weak}$ long-range interactions (see Fig.~\ref{Fig:2}(a)) where we have observed a maximum close to the threshold at least for sufficiently large $\alpha\geq 1.5$. We compare our results obtained from FGS also with the ones obtained from DMRG results. Here, we find that DMRG always predicts a lower ground state energy. The discrepancy of the two methods is even more striking in the entanglement entropy visible in Fig.~\ref{Fig:5}(b). Here, while we still observe very good agreement for $\alpha=0.75$ we found clear deviations for $\alpha=0.3$. The DMRG results predict tendentially a larger entanglement entropy than the FGS. This is an indicator that FGS are less well-suited for the description of the TFIM for very small $\alpha$, i.e. very strong long-range interactions.

\subsubsection{Violations to the area law\label{alpha_less_1}}
\begin{figure}[t]
    \centering
    \includegraphics[width=1\columnwidth]{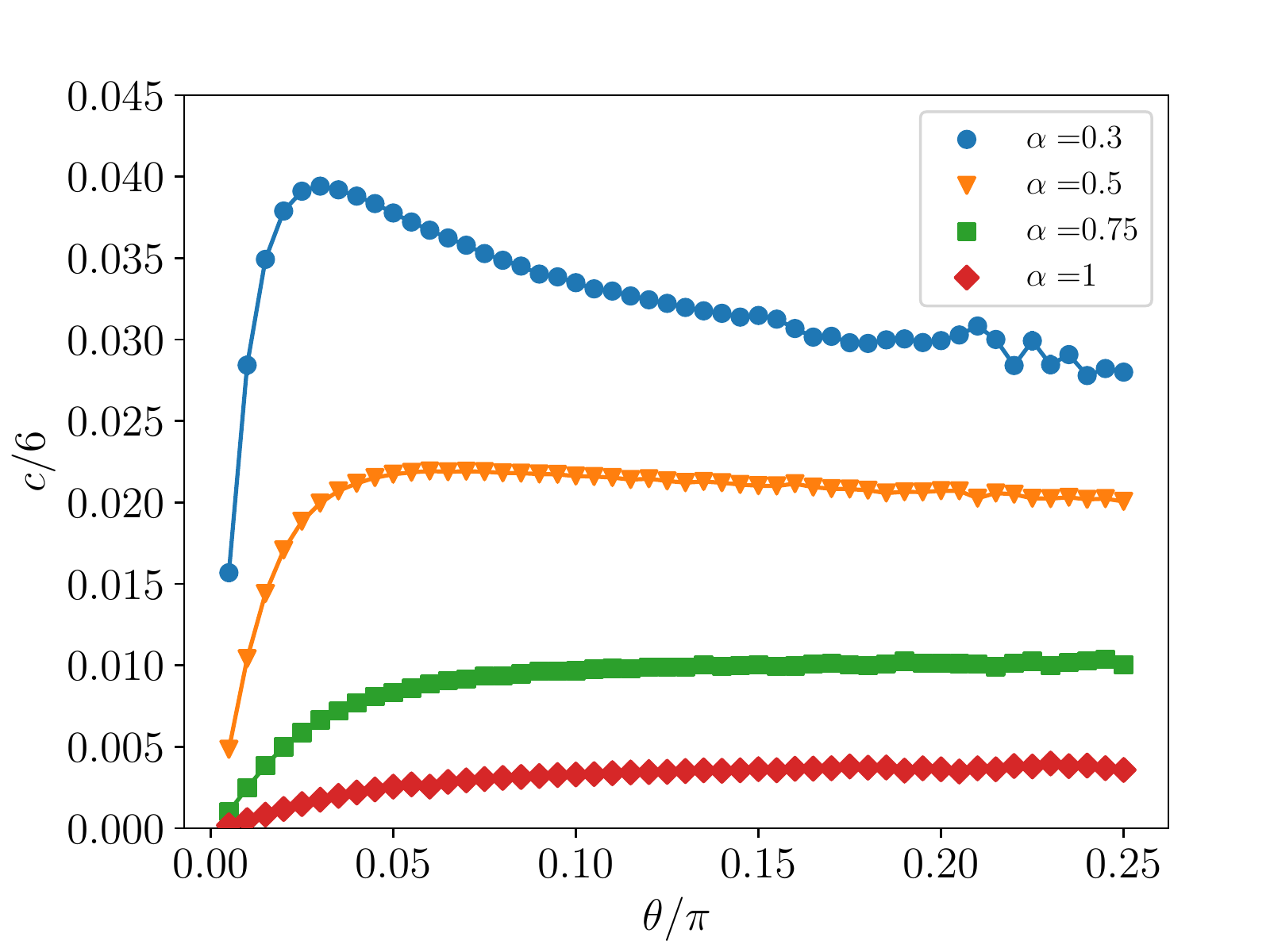}\label{fig:1c}
    \caption{\label{Fig:6} Violations to the area law: The effective central charge $c$ [Eq.~\eqref{eq:EE}] calculated from finite scaling of system sizes $N\in\{40,50,\dots,100\}$ for 50 different values deep in the gapped region $\theta\in(0,\pi/4)$ for the GHF ITE algorithm. Error bars for the standard deviation  are also included, but too small to be visible. }
\end{figure}

We will now analyze the scaling of the entanglement entropy with the system size. For this we calculate the entanglement entropy for various parameters $\theta$ and $\alpha$ and for different numbers of spins $N\in\{40,50,\dots,100\}$.
We then fit the coefficients $c$ and $B$ using Eq.~\eqref{eq:EE} to the obtained values of the entanglement entropy. The obtained values of $c$ are shown in Fig.~\ref{Fig:6}. At this point we remark that the effective central charge $c$ is calculated far away from the threshold in a phase with a non-vanishing energy gap ~\cite{koffel2012entanglement}.

For the values $\alpha<1$, we find $c=0$ only at $\theta=0$. For increasing $\theta$ we find a sharp increase of $c$. For $\alpha=0.3$ and $\alpha=0.5$ we find a maximum and then a decrease again for larger values of $\theta$. A qualitatively similar behavior has also been observed in Ref.~\cite{koffel2012entanglement}. This has been seen as a violation to the area law since this logarithmic divergence does not originate from a closing gap in the spectrum of the system~\cite{koffel2012entanglement}. We therefore conclude that the FGS are able to predict this feature, although the quantitative values deviate from the ones obtained from DMRG results.

%-------------------------------------------------------

%-------------------------------------------------------
%-------------------------------------------------------
%-------------------------------------------------------
\section{Summary and outlook \label{summary}}
%-------------------------------------------------------
This work presents an extensive study of the AFM long-range TFIM in both the weak and strong long-range regime using generalized Hartree Fock theory, a  mean-field method with low computational cost. We validate our results by comparing the computed energy and entanglement entropy to DMRG. We plot the phase diagram and provide estimates for the location of the critical point of the second order phase transition through finite-size scaling for $\alpha\in[1,3]$ and find that they are in excellent agreement with both LCE calculations of Ref.~\cite{fey2019quantum} and DMRG simulations of Refs.~\cite{koffel2012entanglement,vodola2015long}. At the critical point, we compute the central charge $c$ of the underlying conformal field theory for $\alpha\in\{0.3,0.5,1\}$, and find $c>1/2$ for all values of $\alpha$. In the strong long-range regime we still found qualitative agreement between FGS and DMRG calculations. Hereby we found larger quantitative deviations for smaller values of $\alpha$. Remarkably, GHF can predict the logarithmic violations to the area law in the AFM-TFIM which has previously been studied with DRMG. Based on these findings, we conclude that FGS provide a numerically inexpensive alternative to study the AFM long-range TFIM and that our results are in good agreement with DMRG, the current state-of-the-art numerical method for one-dimensional lattice systems.

All simulations were carried out using a standard laptop computer. Since the dimensionality of the system only appears in the Hamiltonian elements $h_{pq}$ and $J_{pq}$, it is straightforward to apply FGS to the two- and three-dimensional TFIM. Therefore, it would be interesting to compare FGS simulations with methods that can be applied to the two-dimensional AFM-TFIM~ \cite{fey2019quantum}.
Moreover, while we have focused on the AFM regime, FGS can readily be applied to the ferromagnetic regime $\theta\in(-\pi,0)$. In this work we have focused on the entanglement entropy, however, pair correlation functions and the entanglement spectrum can be extracted from the covariance matrix as well.  FGS can also be used to study dynamics under the evolution of the TFIM, with equations of motion similar to Eq.~\eqref{is25} \cite{kraus2010generalized,shi2018variational}. In particular, studying the dynamics of the entropy after a quench would offer the possibility to verify the breaking of conformal symmetry in the regime $\alpha <1$ \cite{schachenmayer2013entanglement}. From a numerical standpoint, more efficient calculations of the central quantities such as Eqs.~\eqref{is27} could lead to dramatic computational speedups. As a possible pathway, it would be interesting to see if sum-identities for Pfaffians such as provided in Refs.~\cite{ishikawa1995minor,ishikawa2000minor,ishikawa2006applications} could be applied to the TFIM Hamiltonian. Finally, one could study if different spin-to-fermion mappings~\cite{bravyi2002fermionic,tranter2015bravyi,Jiang2020optimalfermionto}, each resulting in a different form of $H$ when expressed in fermionic operators, have an effect on the FGS simulations. 
%-------------------------------------------------------
%-------------------------------------------------------
%-------------------------------------------------------
\begin{acknowledgments}
The authors thank Kai Phillip Schmidt for providing the data for the LCE calculations from Ref.\cite{fey2019quantum} and Luca Tagliacozzo for providing the DMRG data from Ref.~\cite{koffel2012entanglement}. The authors also thank Giovanna Morigi for insightful discussions. M.K. thanks Miguel \'Angel Mart\'in-Delgado and Frank Wilhelm-Mauch for helpful discussions and support. S.B.J. acknowledges support from the Research Centers of the
Deutsche Forschungsgemeinschaft (DFG): Projects A4 and
A5 in SFB/Transregio 185: “OSCAR.”
\end{acknowledgments}
%-----------------------------------------
%-----------------------------------------
%-----------------------------------------
%\bibliography{main}% Produces the bibliography via BibTeX.

%-----------------------------------------
%-----------------------------------------
%-----------------------------------------
\onecolumngrid
\appendix

%-----------------------------------------
%-----------------------------------------
%-----------------------------------------
\section{Derivation of the equations of motion for the ITE algorithm\label{derivation_ITE}}
In this section, we will derive Eq.~\eqref{is25} which describes the imaginary-time evolution of a FGS. Eq.~\eqref{is25} was also shown in Ref.~\cite{kraus2010generalized} for fourth-order polynomials of fermionic opertors and for an even more general case in Ref.~\cite{shi2018variational}.

We start by writing down the imaginary-time time evolution for the pure FGS $\hat \rho_{\mathrm{GS}}=\ket{\Psi_{\text{GS}}}\bra{\Psi_{\text{GS}}}$ determined by
\begin{align}
\frac{d}{d\tau} \ket{\Psi_{\text{GS}}} = -\left(\hat H-\braket{\Psi_{\text{GS}}|\hat H|\Psi_{\text{GS}}}\right)\ket{\Psi_{\text{GS}}}.\label{a27}
\end{align}
The pure FGS can be generated by a Gaussian transformation 
\begin{align}
\ket{\Psi_{\text{GS}}} =& \hat U_{\text{GS}} \ket{\text{vac}},\label{a1}
\end{align}
where $\ket{\text{vac}}$ denotes the fermionic vacuum and 
\begin{align}
\hat U_{\text{GS}}(\boldsymbol\xi) =& e^{\frac{i}{4}\mathbf{\hat a}^T\boldsymbol\xi\mathbf{\hat a}}\label{a2}
\end{align}
describes the generator of a pure FGS~\cite{bravyi2004lagrangian}. Here, $\boldsymbol\xi$ denotes a $(2n\times 2n)$ anti-symmetric and Hermitian matrix (the matrix elements $\xi_{kl}=-\xi_{lk}$ are purely imaginary). To calculate the covariance matrix $\boldsymbol\Gamma$ we use
\begin{align}
    \boldsymbol\Gamma = -\mathbf U_\xi \boldsymbol \Upsilon \mathbf U_\xi^T,\label{a21}
\end{align}
where  
\begin{align}
    \boldsymbol\Upsilon = \bigoplus_{p=1}^N\begin{pmatrix}
    0&1\\-1&0
    \end{pmatrix}.\label{a22}
\end{align}
is the covariance of the vacuum state and where we employed the transformation
\begin{align}
    \hat U_{\text{GS}}^\dag(\boldsymbol\xi) \hat{\mathbf a} \hat U_{\text{GS}}(\boldsymbol\xi) = \mathbf U_\xi\hat{\mathbf a},\label{a4}
\end{align}
with
\begin{align}
    \mathbf U_{\xi} = e^{i\boldsymbol\xi}.\label{a3}
\end{align}

Now, the idea is that we derive from Eq.~\eqref{a27} a differential equation for ${\bf U}_{\xi}$ which can then be used to calculate $\boldsymbol\Gamma $ using Eq.~\eqref{a21}. For this we treat the left- and right-hand side of Eq.~\eqref{a27} separately and rewrite it as 
\begin{align}
    \hat{U}_{\mathrm{GS}}\hat{L}\ket{\mathrm{vac}}=\hat{U}_{\mathrm{GS}}\hat{R}\ket{\mathrm{vac}}.\label{a90}
\end{align}
We then expand the operators $\hat{L}$ and $\hat{R}$ up to second order in terms of normal-ordered monomials of the Majorana operators and apply them to the vacuum state. 

\paragraph{Left-hand side of Eq.~\eqref{a27}:}
The operator $\hat{L}$ is defined as
\begin{align}
    \hat{L}=\hat{U}_{\mathrm{GS}}^\dag\left(\frac{d\hat U_{\text{GS}}(\boldsymbol\xi)}{d\tau}\right).\label{a60}
\end{align}
The derivative of the unitary transformation is given by
\begin{align}
\frac{d\hat U_{\text{GS}}(\boldsymbol\xi)}{d\tau} 
=&U_{\text{GS}}(\boldsymbol\xi)\left[\frac{i}{4}\hat{\mathbf a}^T\mathbf U_\xi^T\frac{d\mathbf U_\xi}{d\tau}\hat{\mathbf a}\right].
\label{a26}
\end{align}
Here, we have used Eq.~\eqref{a2}, the identity \cite{wilcox1967exponential}
\begin{align}
\frac{de^{\hat J(\tau)} }{d\tau} =& \int_0^1du e^{(1-u)\hat J(\tau)}\left(d_\tau \hat J(\tau)\right)e^{u\hat J(\tau)},\label{a25}
\end{align}
and the orthogonality property $\mathbf U_\xi\mathbf U_\xi^T=\bf 1$.
For the normal-ordered expression we therefore find
\begin{align}
\hat{L}=\hat{L}_0+\hat{L}_2,
\end{align}
with
\begin{align}
\hat{L}_0=&\frac{i}{4}\text{tr}\left[\frac{d\mathbf U_\xi}{d\tau} \mathbf U_\xi^T \boldsymbol\Gamma\right],\label{a44}\\
\hat{L}_2=&\frac{1}{4}:\hat{\mathbf a}^T \mathbf U_\xi^T\frac{d\mathbf U_\xi}{d\tau}\hat{\mathbf a}:,\label{a45}
\end{align}
where $:\hat{A}:$ denotes the elementwise normal-ordering of $\hat{A}$. 
\paragraph{Right-hand side of Eq.~\eqref{a21}:}
The definition of $\hat{R}$ is
\begin{align}
\hat{R}=-\hat{U}_{\mathrm{GS}}^\dag\left(\hat{ H}-\bra{\Psi_{\mathrm{GS}}}\hat{H}\ket{\Psi_{\mathrm{GS}}}\right)\hat{U}_{\text{GS}}.\label{a61}
\end{align}
We can now use a modification of Wicks theorem to calculate
\begin{align}
    \hat{U}_{\mathrm{GS}}^\dag\hat{H}\hat{U}_{\mathrm{GS}}=\bra{\Psi_{\mathrm{GS}}}\hat{H}\ket{\Psi_{\mathrm{GS}}}+\frac{i}{4}:\hat{\mathbf a}^T \mathbf U_\xi^T{\bf H}^{(\mathrm{mf})}{\bf U}_\xi{\bf a}:+\tilde{Q},\label{a95}
\end{align}
where $\tilde{Q}$ collects all normally ordered monomials of quartic order or higher. We derive this expression in Appendix~\ref{App:Wicktheorem}. Inserting Eq.~\eqref{a95} into Eq.~\eqref{a61}, we find for $\hat{R}$ the following expression
\begin{align}
    \hat{R}=\hat{R}_2-\tilde{Q},
\end{align}
with
\begin{align}
    \hat{R}_2=-\frac{i}{4}:\hat{\mathbf a}^T \mathbf U_\xi^T{\bf H}^{(\mathrm{mf})}{\bf U}_\xi{\bf a}:.
\end{align} 

\paragraph{Comparing left- and right-hand side:}
We now require to match $\hat{L}\ket{\text{vac}}$ and $\hat{R}\ket{\text{vac}}$ up to second order. This is a consequence of our restriction to FGS. Therefore, we find two equations
\begin{align}
    \hat{L}_0\ket{\text{vac}}=&0,\label{eq:L0}\\
    \hat{L}_2\ket{\text{vac}}=&\hat{R}_2\ket{\text{vac}},\label{eq:L2}
\end{align}
from which we wish to derive the equations of motion of the ITE. We first consider Eq.~\eqref{eq:L2},
\begin{align}
    :\hat{\mathbf a}^T \mathbf U_\xi^T\frac{d\mathbf U_\xi}{d\tau}\hat{\mathbf a}:\ket{\text{vac}} =& -i:\hat {\mathbf a}^T\mathbf U_\xi^T \mathbf H^{(\text{mf})} \mathbf U_\xi\hat {\mathbf a}:\ket{\text{vac}}.\label{a47}
\end{align}
For any normal-ordered polynomial of fermionic operators applied to the vacuum state, the only terms that do not vanish are polynomials which exclusively contain fermionic creation operators. Therefore, we define the vector
\begin{align}
    \mathbf r = \hat{\mathbf c}^\dag\otimes\begin{pmatrix}1\\i\end{pmatrix},\label{a92}
\end{align}
where $\hat{\mathbf c}^\dag=(\hat c_1^\dag,\dots,\hat c_N^\dag)$, and rewrite Eq.~\eqref{a47} in terms of fermionic creation and annihilation operators, which leads to
\begin{align}
    \hat{\mathbf r}^T \mathbf U_\xi^T\frac{d\mathbf U_\xi}{d\tau}\hat{\mathbf r}\ket{\text{vac}} =& -i\hat {\mathbf r}^T\mathbf U_\xi^T \mathbf H^{(\text{mf})} \mathbf U_\xi\hat {\mathbf r}\ket{\text{vac}},\label{a93}
\end{align}
where the normal-ordering ``$:\ :$'' may now be dropped. We would now like to compare the matrices of Eq.~\eqref{a93}. However, before doing so, we  need to take into account the symmetry operations which leave the operator $\hat{\mathbf r}$ invariant. For this, we first rewrite $\boldsymbol\Upsilon$ defined in Eq.~\eqref{a22} as  $\boldsymbol\Upsilon = \bf 1_N\otimes\left(\begin{smallmatrix}0&1\\-1&0\end{smallmatrix}\right)$. Thus, the symmetry operations on the operators are given by $-i\boldsymbol\Upsilon\hat{\mathbf r} = \hat{\mathbf r}$ and $i\hat{\mathbf r}^T\boldsymbol\Upsilon = \hat{\mathbf r}^T$. The real-valued skew-symmetric solution for $\frac{d\mathbf U_\xi}{d\tau}$ which satisfies Eq.~\eqref{eq:L0} is then given by
\begin{align}
\frac{d\mathbf U_\xi}{d\tau} =-\frac{1}{2}\boldsymbol\Gamma\mathbf H^{(\text{mf})} \mathbf U_\xi -\frac{1}{2}\mathbf H^{(\text{mf})}\mathbf U_\xi\boldsymbol\Upsilon \label{a48}.
\end{align}
For the derivative of the covariance matrix $\boldsymbol{\Gamma}$ we find then
\begin{align}
\frac{d\boldsymbol\Gamma}{d\tau}  =& -\frac{d\mathbf U_\xi}{d\tau}\boldsymbol\Upsilon  \mathbf U_\xi^T - \mathbf U_\xi \boldsymbol\Upsilon \frac{d\mathbf U_\xi^T}{d\tau}\nonumber\\
&=-\mathbf H^{(\text{mf})} - \boldsymbol\Gamma\mathbf H^{(\text{mf})}\boldsymbol\Gamma,\label{a51}
\end{align}
which is identical to Eq.~\eqref{is25} using $\boldsymbol{\Gamma}^2=-{\bf 1}$.

\section{Best quadratic approximation\label{App:Wicktheorem}}
In this section we will show Eq.~\eqref{a95}, which can be derived using Wick's theorem. In particular, we will derive this formula for an arbitrary Hamiltonian which is a sum of even products of Majorana operators. By the application of normal-ordering onto a polynomial $\hat p(k)$ of order $k$ we understand the sum of normal-ordered monomials $\hat m(l)$ of order $l\leq k$, in other words $:\hat p(k):=\sum_{l=0}^k:\hat m(l):$. Therefore, it is sufficient to show the relation~\eqref{a95} for arbitrary even products of Majorana operators. Without loss of generality we number the Majorana operators from $1,...,2n$ with $n\in\mathbb{N}$.
Using normal-ordering and Wicks theorem we can write the product $\hat{A}=\hat{a}_1\hat{a}_2\dots\hat{a}_{2n}$ (a monomial of Majorana operators) in the following way
\begin{align}
\hat{A}=\bra{\mathrm{vac}}\hat{A}\ket{\mathrm{vac}}+\sum_{i<j}(-1)^{i+j+1}\bra{\mathrm{vac}}\hat{A}_{\hat{i},\hat{j}}\ket{\mathrm{vac}}:\hat{a}_i\hat{a}_j:+\hat{Q},
\end{align}
where $\hat{Q}$ collects all normal-ordered monomials which are at least quartic in the Majorana operators. We furthermore introduced the reduced product $\hat{A}_{\hat{i},\hat{j}}$ which emerges from $\hat{A}$ by removing the operators $\hat{a}_i$ and $\hat{a}_j$.
If we transform the vacuum state using Eq.~\eqref{a1}, this expansion needs to be modified according to
\begin{align}
\hat{U}_{\mathrm{GS}}\hat{A}\hat{U}_{\mathrm{GS}}^\dag=\bra{\Psi_{\text{GS}}}\hat{A}\ket{\Psi_{\text{GS}}}+\sum_{i<j}(-1)^{i+j+1}\bra{\Psi_{\text{GS}}}\hat{A}_{\hat{i},\hat{j}}\ket{\Psi_{\text{GS}}}:\hat{U}_{\mathrm{GS}}\hat{a}_i\hat{a}_j\hat{U}_{\mathrm{GS}}^\dag:+\tilde{Q},
\end{align}
where $\tilde{Q}$ collects terms of order four and higher. Denoting ${\bf A}=\left.\boldsymbol{\Gamma}\right|_{1,2,\dots,2n}$ and using Eq.~\eqref{iis13}, we can rewrite the above equation and find
\begin{align}
\hat{a}_1\hat{a}_2\dots\hat{a}_{2n}=(-i)^{n}\text{Pf}({\bf A})+\sum_{i<j}(-i)^{n-1}(-1)^{i+j+1}\text{Pf}({\bf A}_{\hat{i}\hat{j}}):\hat{U}_{\mathrm{GS}}\hat{a}_i\hat{a}_j\hat{U}_{\mathrm{GS}}^\dag:+\tilde{Q}.
\end{align}
Here, we have introduced the submatrices ${\bf A}_{\hat{i}\hat{j}}$ which emerge from ${\bf A}$ by canceling the $i$th and $j$th rows and columns. Using 
\begin{align}
\frac{\partial \text{Pf}({\bf A})}{\partial \Gamma_{ij}}=\frac{(-1)^{i+j+1}}{2}\text{Pf}({\bf A}_{\hat{i}\hat{j}}),\label{eq:derivativePfaffian}
\end{align}
we obtain the expression
\begin{align}
\hat{A}=&(-i)^{n}\text{Pf}({\bf A})+i\sum_{i,j}(-i)^{n}\frac{\partial \text{Pf}({\bf A})}{\partial \Gamma_{ij}}:\hat{U}_{\mathrm{GS}}\hat{a}_i\hat{a}_j\hat{U}_{\mathrm{GS}}^\dag:+\tilde{Q}\nonumber\\
=&\bra{\Psi_{\text{GS}}}\hat{A}\ket{\Psi_{\text{GS}}}+\frac{i}{4}:\hat{\mathbf a}^T \mathbf U_\xi^T{\bf A}^{(\mathrm{mf})}{\bf U}_\xi{\bf a}:+\tilde{Q},\label{eq:norm_exp}\end{align}
with matrix entries
\begin{align}
A_{ij}^{(\mathrm{mf})}=4\frac{\partial\bra{\Psi_{\text{GS}}}\hat{A}\ket{\Psi_{\text{GS}}}}{\partial \Gamma_{ij}}.
\end{align}
We want to remark that Eq.~\eqref{eq:derivativePfaffian} can also be used to efficiently calculate the derivative Eq.~\eqref{eq:mf} of Eq.~\eqref{is27}.

%-----------------------------------------
%-----------------------------------------
%-----------------------------------------
%-----------------------------------------
%-----------------------------------------
%-----------------------------------------
%-----------------------------------------

\end{document}